# System Analysis Modeling and Intermodal Transportation for Commercial Spent Nuclear Fuel

H.R. Gadey[a], M. Nutt[a], P. Jensen[a], R. Howard[a], R. Joseph[b], L. Vander Wal[c], R. Cumberland[d]

[a] Pacific Northwest National Laboratory, 902 Battelle Blvd, Richland, WA 99354 (harish.gadey@pnnl.gov)
[b] Idaho National Laboratory, 1955 N Fremont Ave, Idaho Falls, ID 83415
[c] Argonne National Laboratory, 9700 S Cass Ave, Lemont, IL 60439
[d] Oak Ridge National Laboratory, 5200, 1 Bethel Valley Rd, Oak Ridge, TN 37830

## INTRODUCTION

The United States Department of Energy (DOE) Office of Nuclear Energy is applying knowledge and understanding in the areas of systems level engineering, analysis, and decision making to better inform the waste management pathways for U.S. spent nuclear fuel (SNF). Currently there are 93 operating and 23 shutdown commercial nuclear reactors in the United States. SNF at most of these locations is being stored in spent fuel pools, dry storage, or both [1, 2]. This paper initially goes over the basics of the agent-based simulation tool known as the Next Generation System Analysis Model (NGSAM) that has the capability to model interaction and movements of individual components or groups like casks, railcars etc. The next section covers some of the methods implemented in the Java Transportation Operations Model (JTOM) to facilitate the movement of assemblies, casks, railcars etc. and the various intermodal transfer options. This section also covers the current cask loading and intermodal transfer times implemented in NGSAM. The last section goes over the proposed values for intermodal transportation, and cask transfer/loading operations.

## NEXT GENERATION SYSTEM ANALYSIS MODEL

NGSAM is an agent-based simulation toolkit that is used for system level simulation and analysis of the U.S. SNF inventory [3]. This tool was developed as part of a collaborative effort between Oak Ridge National Laboratory (ORNL), Sandia National Laboratories (SNL), and Argonne National Laboratory (ANL). NGSAM was based on the process analysis tool used by the Department of Defense and the Federal Emergency Management Agency to accomplish system level logistical models. The analyst using NGSAM has the freedom to define multiple factors like the number of storage facilities, capacity at each facility, transportation schedules, shipment rates, and other reactor operating conditions. The primary purpose of NGSAM is to provide the system analyst with the tools to model the integrated waste management system. Some of the system analyst goals that can be accomplished using NGSAM include:

- Gathering information regarding waste management alternatives and considerations.
- Studying the impact analysis of storage choices.
- Incorporating a wide array of aspects such as repackaging needs, repository emplacement capacity, storage, and transportation alternatives, etc. in a single scenario.
- Developing an integrated approach with emphasis on flexibility.

Data required for performing the NGSAM analysis comes from two different sources; the Unified Database (UDB) which is used with the Used Nuclear Fuel Storage, Transportation, & Disposal Analysis Resource and Data System (UNFST&DARDS) and the scenario-specific data. The UDB provides fuel inventory, canister and cask specific data, while the scenario specific data is information that is entered by the analyst for their respective use case.

One of the main features of NGSAM is the transportation analysis of SNF from reactor sites to various locations including hypothetical monitored geological repositories (MGRs), repackaging facilities, and interim storage facilities (ISFs). These transportation operations are carried out using the ORNL's JTOM. ORNL modified the transportation operations model (TOM) to enable integration with the NGSAM front end [4]. JTOM requires data from several sources to run the transportation models; the assembly and canister specific information originates from the UDB, while the transportation routing information is generated using the Stakeholder Tool for Assessing Radioactive Transportation (START) outputs. JTOM functions in the capacity of a scheduler and NGSAM uses JTOM to schedule movement of SNF from various sites. For example, in a run, a rail consist can originate from a fleet maintenance facility (FMF) and move to an intended pickup location. This is followed by routing to the drop off location, and finally back to the FMF. Fig. 1 shows an example of a transportation run between an ISF and a MGR. In addition, JTOM is also used by NGSAM for performing certain transportation cost estimations.

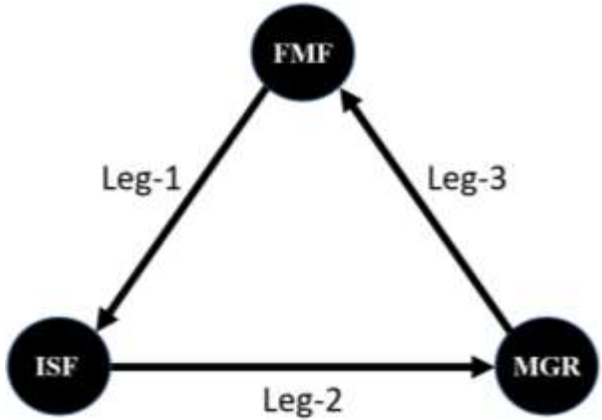

Fig. 1. A sample transportation itinerary in JTOM.

**METHODS AND CURRENT IMPLEMENTATION**

While executing a scenario, several JTOM methods are called by NGSAM. Some of the important methods from a loading and intermodal transportation standpoint are:

- Set Site Handling Time: the site handling time method deals with the amount of time it takes to load an empty transportation cask with SNF.
- Set Site Transfer Time: the site transfer time is the amount of time it takes to place loaded or empty casks on transportation assets like barges, rail, or heavy haul trucks; or the time it takes to unload them from these assets.
- Set Mode Transfer Time: mode transfer time is the amount of time it takes to transfer an empty or loaded cask from one transportation mode to the other. For example, from a barge to rail or from rail to heavy haul truck.
- Set Operating Hours: operating hours method sets the number of hours per day a site is operational.
- Add To Rail Route Stage: sometimes a direct rail route to a reactor site is not readily available, under those circumstances, the "add to rail route" method is called to add a secondary stage of the route either via barge, heavy haul truck, or the combination of the two. The use of multiple transportation modes requires the use of a transload site (TLS), a location where loaded or empty casks are transferred from one mode to the other.

Data for the transfer times, site hours, and add to rail route methods are often stored in tables within NGSAM or CSV files that are imported while running a scenario. It is also interesting to note that some of the methods like site transfer times or mode transfer times have data stored for empty as well as loaded casks since empty casks are also required to be transferred on site from an external location prior to loading.

Rail is generally the preferred mode of transportation from the reactor to any off-site location considering the mass, size, and quantity of the casks shipped in each run. It must be noted that barges can also carry similar if not a greater number of casks in a single run, however rail proves to be effective in connecting more geographical regions across the U.S. compared to navigable waterways. Under some circumstances, the reactor site might not be readily accessible using a rail line. Under those scenarios, a barge or heavy haul truck shipment is used to transfer casks from a reactor site to an intermediate site (TLS) where the casks are transferred from barge/heavy haul truck to rail. It is worth mentioning that casks can also be transferred from rail to the reactor site via a barge or heavy haul truck. This process is generally referred to as intermodal transportation where more than one mode of transportation is used to transfer casks from source to destination. Within each transportation mode, there are several sub modes, for example casks can be loaded on a rail directly, or under some cases a staging area (SA) is required to be set and a site transporter is used to transfer casks between the independent spent fuel storage installation (ISFSI) and the SA (protected area/ on-site transfer). The sub mode likely to be used for a particular reactor site depends on the site-specific infrastructure availability and operability. In addition, the following factors may also play a role in identifying the most suitable transportation option:

- Distance to and cost of using an off-site transfer location.
- Rail length and ease of access to spent fuel pools (SFP) or ISFSI.
- Distance between SFP or ISFSI and the track/barge slip.
- Cask receipt area capacity and infrastructure to accommodate a railcar.

In the release version of NGSAM (version 2.3.11.0), scenarios are executed assuming 16 hours to load and 16 hours to unload SNF into or from a transportation cask/overpack. This data is generally used by the set site handling time method. Table 1 shows the load and unload times for overpacks using various transportation modes. This is the data used by the set site transfer time method.

TABLE I. Load and Unload Times for Various Transportation Assets

| Mode | Load Hours per Asset | Unload Hours per Asset |
|---|---|---|
| Rail | 4 | 4 |
| Heavy Haul Truck | 4 | 4 |
| Barge | 5 | 5 |

Table 2 illustrates some of the intermodal transfer times that are utilized by the set mode transfer time method.

TABLE 2. Transfer Times for Casks/Overpacks between Various Modes

| From Mode | To Mode | Loaded Cask (Y/N) | Transfer Time (hours) |
|---|---|---|---|
| Rail | Barge | No | 4 |
| Barge | Rail | Yes | 4 |
| Rail | Heavy Haul Truck | No | 3 |
| Heavy Haul Truck | Rail | Yes | 4 |

In addition to the transfer times noted in Table 2, eight and sixteen additional hours are required to adequately equip the rail consist (including buffer and escort railcars) and barge respectively prior to shipping the overpacks.

Currently the "set site handling time" method in NGSAM uses cask/overpack loading times that are identical irrespective of reactor type. If loading SNF directly from an SFP, generally boiling water reactor (BWR) SNF takes a little longer compared to pressurized water reactor (PWR) SNF to load since a greater number of BWR fuel assemblies can be accommodated in a cask compared to PWR fuel assemblies. Also, at most reactor sites ISFSIs are already present, meaning fuel is already stored in dry storage. It must also be noted that moving casks or overpacks from dry storage takes significantly less time compared to moving from the SFP. Thus, an area for refinement in NGSAM would be to identify "set site handling time" values based on particular cask types and/or reactor types and loading location (from SFP or ISFSI) as data becomes available. The following sections describe some work recently conducted to provide additional definition to these time components.

**UPDATED NGSAM TRANSFER TIMES**

This work discusses steps from when the casks arrive at the TLS facility to when the casks are loaded back on railcars. The approximate time for most of the steps is estimated based on data from previous DOE-sponsored studies [5,6]. Three scenarios were explored in this work: barge to TLS, barge to TLS with a short distance heavy haul truck, and heavy haul truck to TLS. In the second option, a heavy haul truck is initially used for a short distance to move the casks between the reactor site and the barge. This is followed by transporting the casks to the TLS via a barge. The individual steps, transfer times (BWR SFP, PWR SFP, and ISFSI), and the appropriate JTOM methods (if available) are identified for each step and transportation mode. Several assumptions were made in this work including:
- The availability of a SA on site in the protected area to load and unload the casks.
- Five casks are assumed per rail and barge shipment.
- The use of two heavy haul trucks for the heavy haul truck to TLS option is considered.
- The use of roll-on roll-off procedure for cask movements using a barge.
- Availability of a site transporter to move casks between the ISFSI, SFP, or SA.

**Barge to Transload Site**

This section covers the various steps involved in transferring a cask from the TLS to the reactor site and back via a barge (Table 3).

TABLE 3. Transload Operations, Transfer Times, and JTOM Methods for a Barge to TLS Operation

| Steps | Transfer time (h) | | | JTOM Methods |
|---|---|---|---|---|
| | PWR SFP | BWR SFP | ISFSI | |
| 1. Rail consist arrives at TLS | - | - | - | - |
| 2. Casks loaded on barge | 24 | 24 | 24 | Set Mode Transfer Time |
| 3. Travel to reactor site | Site Specific | | | Add to Rail Route |
| 4. Casks unloaded, transfer to SA | 12 | 12 | 12 | Set Site Transfer Time |
| 5. Cask to SFP/ISFSI [a] | 17.6 | 17.6 | | |
| 6. Cask loaded [a] | 95 | 104.9 | 48 | Set Site Handling Time |
| 7. Loaded cask to SA [a] | 23.8 | 23.8 | | |
| 8. Loaded casks from SA to barge | 12 | 12 | 12 | Set Site Transfer Time |
| 9. Travel to TLS | Site Specific | | | Add to Rail Route |
| 10. Casks loaded on rail | 144 | 144 | 144 | Set Mode Transfer Time |
| 11. Configure buffer and escort cars | 24 | 24 | 24 | |

[a] These values indicate time per individual cask. To estimate the total time, these values are multiplied by the total number of overpacks in the campaign (5).

## Barge to Transload Site with a Short Distance Heavy Haul Truck

The various steps involved in the barge to TLS cask transfer operation with a short-distance heavy haul truck is illustrated in table 4.

TABLE 4. Transload Operations, Transfer Times, and JTOM Methods for a Barge to TLS using Short Haul Truck Operation

| Steps | Transfer time (h) | | | JTOM Methods |
|---|---|---|---|---|
| | PWR SFP | BWR SFP | ISFSI | |
| 1. Rail consist arrives at TLS | - | - | - | - |
| 2. Casks loaded on barge | 24 | 24 | 24 | Set Mode Transfer Time |
| 3. Travel to site | Site Specific | | | Add to Rail Route |
| 4. Casks transferred from barge to SA via truck | 24 | 24 | 24 | Multiple Methods* |
| 5. Cask to SFP/ISFSI [a] | 17.6 | 17.6 | | |
| 6. Cask loaded [a] | 95 | 104.9 | 48 | Set Site Handling Time |
| 7. Loaded cask to SA [a] | 23.8 | 23.8 | | |
| 8. Loaded casks from SA to barge via truck | 24 | 24 | 24 | Multiple Methods* |
| 9. Travel to TLS | Site Specific | | | Add to Rail Route |
| 10. Casks loaded on rail | 144 | 144 | 144 | Set Mode Transfer Time |
| 11. Configure buffer and escort cars | 24 | 24 | 24 | |

[a] These values indicate time per individual cask/overpack. To estimate the total time, these values are multiplied by the total number of overpacks in the campaign (5).
* Set Site Transfer Time, Add to Rail Route, and Set Mode Transfer Time methods are intended to be used in these steps.

## Heavy Haul Truck to Transload Site

The third transportation option covered in this work is transporting casks to the TLS via heavy haul truck. Table 5 illustrates the various steps involved in this process.

TABLE 5. Transload Operations, Transfer Times, and JTOM Methods for a Heavy Haul Truck to TLS Operation

| Steps | Transfer time (h) | | | JTOM Methods |
|---|---|---|---|---|
| | PWR SFP | BWR SFP | ISFSI | |
| 1. Rail consist arrives at TLS | - | - | - | - |
| 2. Casks loaded on truck | 24 | 24 | 24 | Set Mode Transfer Time |
| 3. Travel to site | Site Specific | | | Add to Rail Route |
| 4. Cask to SFP/ISFSI [a] | 17.6 | 17.6 | | |
| 5. Cask loaded [a] | 95 | 104.9 | 72 | Set Site Handling Time |
| 6. Cask transferred from SFP/ ISFSI [a] | 23.8 | 23.8 | | |
| 7. Travel to TLS | Site Specific | | | Add to Rail Route |
| 8. Casks loaded on rail | 24 | 24 | 24 | Set Mode Transfer Time |
| 9. Configure buffer and escort cars | 24 | 24 | 24 | |

[a] These values indicate time per individual cask/overpack. To estimate the total time, these values are multiplied by the total number of overpacks in the campaign (5).

## OBSERVATIONS AND FUTURE WORK

In this work, the high-level goals, and functionalities of NGSAM were covered followed by a discussion of some of the transportation and site operation methods in JTOM. The various steps involved in intermodal transportation followed by the transportation times and JTOM method implementation were explored. As it can be seen in the above tables, JTOM methods don't explicitly exist for some of the steps. In such a scenario, either the functional definition of one of the methods is required to be expanded or a new JTOM method shall be defined to account for a given step. It must also be mentioned that for some of the steps, the existing JTOM methods only partially cover the scope of the step, for example the set site transfer time method only accounts for the time to transfer a cask from a barge or heavy haul truck but not the time to transfer the cask to the SA. Site specific modifications are anticipated to be performed in the future to account for this difference. It is also interesting to note that operating hours are captured in NGSAM scenarios for reactor, ISF, MGR, and repackaging sites only while this information is required to be explicitly defined for the TLS as well. To refine the simulations even further it might also be beneficial to study the TLS infrastructure requirements to accurately estimate the transfer times. During a transportation operation, the distance and speed for a given mode of transportation are generally furnished, based on which the transportation time is calculated. Other factors like traffic, weather, infrastructure operation times (raising and lowering locks for waterways), and human factors can also be explored to arrive at more realistic travel times in the future.

**ENDNOTES**

This is a technical paper that does not take into account contractual limitations or obligations under the Standard Contract for Disposal of Spent Nuclear Fuel and/or High-Level Radioactive Waste (Standard Contract) (10 CFR Part 961). For example, under the provisions of the Standard Contract, spent nuclear fuel in multi-assembly canisters is not an acceptable waste form, absent a mutually agreed to contract amendment. To the extent discussions or recommendations in this paper conflict with the provisions of the Standard Contract, the Standard Contract governs the obligations of the parties, and this paper in no manner supersedes, overrides, or amends the Standard Contract. This paper reflects technical work which could support future decision making by DOE. No inferences should be drawn from this paper regarding future actions by DOE, which are limited both by the terms of the Standard Contract and Congressional appropriations for the Department to fulfill its obligations under the Nuclear Waste Policy Act including licensing and construction of a spent nuclear fuel repository.